# High optical absorption in graphene

S. P. Apell*, G. W. Hanson^ and C. Hägglund^^

*Department of Applied Physics, Chalmers University of Technology
SE-412 96 Göteborg, Sweden.  apell@chalmers.se

^Department of Electrical Engineering, University of Wisconsin-Milwaukee, 3200 N. Cramer St., Milwaukee, Wisconsin 53211, USA.

^^ Department of Chemical Engineering, Stanford University, Stanford, CA 94305-5025, USA.

**Abstract:** A simple analysis is performed for the absorption properties of graphene sandwiched between two media. For a proper choice of media and graphene doping/gating one can approach 50-100% absorption in the GHz-THz range for the one atom thick material. This absorption is controlled by a characteristic chemical potential which depends only on carrier life-time and the indexes of refraction of the dielectric embedding.

**OCIS codes:** (230.0230) Optical devices; (260.0260) Physical optics; (310.0310) Thin films.

## 1. Introduction

Light absorption in thin films has always been a relevant topic in optics, especially from the application point of view. Graphene is in many ways the ultimate thin film, only one atomic layer thick, and has photonic properties of high interest for optoelectronic applications [1]. Noteworthy is that for pristine, unbiased graphene an impressive 2.3% ($\alpha\pi$, where $\alpha = e^2/4\pi\varepsilon_o \hbar c$ is the fine structure constant) of incident visible light is absorbed.

Crucial for the optical performance of small particles and ultra-thin structures is often that relevant (surface) plasmon excitations are available. Recently, an optical switching mechanism using gated graphene, coupling to external radiation through surface plasmon-polaritons rather than directly to incoming photons, has been described [2]. Graphene based sensors are another area of importance, where it has been suggested [3] that graphene ribbons can be used to convert molecular signatures to electrical signals based on graphene plasmons being very sensitive to the molecular analytes one is monitoring. Since for graphene we have the possibility of controlling its "optical" properties with a proper gate voltage and/or doping, which through the chemical potential governs the optical conductivity and thus its spectral signature, a multitude of possible mechanisms for sensing and tunable optics are available over a broad frequency range [4].

If we were to freely tune the optical properties of a film with the thickness of a single layer of graphene, the maximum attainable light absorption would be dictated by the contrast of the surrounding media [5]. This general limit for light absorption in ultrathin films may under favorable conditions (that is for high damping materials with nearly imaginary dielectric constant) be approached by tuning of the film thickness alone, as demonstrated by Driessen et al. [6,7]. However a more widely applicable approach to realize these optimal conditions is to exploit plasmon resonances in nanostructures where metallic elements and other materials adding functionality are combined into nanocomposites. By tuning the geometrical properties and thereby the effective dielectric function of the nanocomposite structure in relation to the dielectric properties of the surrounding media, the impedance of the system can be matched to maximize the absorption [8-11]. Related contrast effects are exploited when making graphene "visible" by placing it on top of silicon wafers or using holes in a metallic screen [12-14]. Here we investigate another line of approach, not invoking surface plasmons or other collective excitations, to realize optimal conditions for light absorption, namely the possibility to tune the optical properties of a single layer of graphene by means of realistic bias voltages and doping levels, and by appropriate choice of the dielectric environment.

In the following section we will see how one can achieve conditions of high absorption in a thin film which is not apparent in the standard optical treatments. The subsequent section deals with the application of this to the case of graphene reaching levels of 50-100% absorption, given certain conditions with respect to doping level and/or gate voltage. We end our paper giving some realistic estimates with respect to crucial parameters for suspended graphene, to yield testable predictions.

## 2. Maximized absorptance in ultrathin films

To put our arguments on a firm basis we will first consider the electrodynamics of a thin film (later our graphene layer) as described in Fig. 1. The film thickness $d$ is much smaller than the wavelength of light in all media. Light is incident at an angle $\theta$ with respect to the surface normal of the thin film, from a dielectric medium of refractive index $n_1 = \sqrt{\varepsilon_1/\varepsilon_o}$ where $\varepsilon_1$ is its dielectric function and $\varepsilon_o$ is the vacuum permeability. The thin film has refractive index $n_2 = \sqrt{\varepsilon_2/\varepsilon_o}$ in terms of its complex dielectric function $\varepsilon_2$ and rests on a substrate of refractive index $n_3 = \sqrt{\varepsilon_3/\varepsilon_o}$. With $q = \omega/c$ being the wave number in free space (frequency $\omega$ and light velocity c) and $p_i/q = \sqrt{(\varepsilon_i/\varepsilon_o) - \sin^2\theta}$ the wave number component perpendicular to the interface for p-polarized light in medium *i*, we can define a surface impedance $Z_i \equiv p_i/\omega\varepsilon_i$. Throughout the paper we address a p-polarized situation and obtain s-polarized results from p-polarized light at normal incidence. Notice that at normal incidence $Z_i \equiv Z_o/n_i$ in terms of the vacuum impedance $Z_o \approx 120\pi\,\Omega$. The reflection coefficient $r_{ij}$ between medium *i* and *j* is now $(Z_j - Z_i)/(Z_j + Z_i)$ and for the situation in Figure 1 we have a total reflection coefficient [15]

$$\tilde{r} = \frac{r_{12} + r_{23}\gamma^2}{1 + r_{12}r_{23}\gamma^2}, \tag{1}$$

and a transmission coefficient

$$\tilde{t} = \frac{(1 + r_{12})(1 + r_{23})\gamma}{1 + r_{12}r_{23}\gamma^2} \tag{2}$$

with $\gamma \equiv e^{ip_2 d}$. In terms of these the reflectance *R* is given by $|\tilde{r}|^2$, the transmittance *T* by $\mathrm{Re}(\frac{Z_1}{Z_3})|\tilde{t}|^2$ and the thin film absorption *A* is then *1-R-T*.

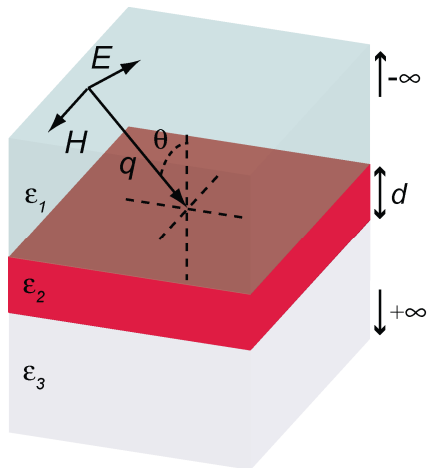

**Figure 1.** A thin film (graphene layer) of width $d$ and complex dielectric function $\varepsilon_2$, embedded between two semi-infinite dielectrics $\varepsilon_1$ and $\varepsilon_3$. A p-polarized electromagnetic wave at an oblique angle $\theta$ is incident from medium *1* (wave number $q$ in vacuum) and has its field vectors oriented as shown. The question addressed in this paper is under what conditions does this structure have an optimum absorption?

Going to the limit of an infinitely thin film ($d \to 0$), defining a two-dimensional conductivity $\sigma_{2D} = \lim_{d \to 0} d\sigma_{3D}$ we can replace the dielectric function $\varepsilon_2$ of the thin film according to

$$\lim_{d \to 0}[-i\omega d\varepsilon_2] = \sigma_{2D}. \tag{3}$$

We can now rewrite Eqs. (1-2) as

$$\widetilde{r} = \frac{Z_3 - Z_1 - Z_1 Z_3 \sigma_{2D}}{Z_3 + Z_1 + Z_1 Z_3 \sigma_{2D}},\tag{4}$$

and

$$\widetilde{t} = \frac{2Z_3}{Z_3 + Z_1 + Z_1 Z_3 \sigma_{2D}},\tag{5}$$

in accordance with the results in [16]. It is the structure of these reflection and transmission coefficients which we will address in what follows. They contain three distinct cases; the surface plasmon-polariton dispersion relation, the traditional thin film limit and the case we pay special attention to in this paper which has the potential of maximizing the absorption in the thin film.

The reflection coefficient of a system can be seen as a general surface electromagnetic response function. In light of this a divergent reflection coefficient corresponds to an intrinsic surface eigen-mode of the response; we can in principle have a reflected field without any incoming one. This property of the reflection coefficient gives thus the dispersion relation $\omega(k)$ for an electromagnetic surface mode of frequency $\omega$ and wave number $k$ parallel to the surface as a solution to (cf. the pole of Eq. (4))

$$\frac{1}{Z_1(\omega,k)} + \frac{1}{Z_3(\omega,k)} + \sigma_{2D}(\omega) = 0,\tag{6}$$

in accordance with the results in [17,18].

Most thin film optical treatments are concerned with the limit of film thickness $d$ being much smaller than the wavelength in the medium. In such a thin electromagnetic system, where the graphene atomic layer is an ultimate limit, an expansion of Equations (4-5) for small $\sigma_{2D} Z_{1,3}$ is the default treatment. For a thin film in vacuum ($Z_1 = Z_3 = Z_o$) and normalizing to the vacuum impedance according to $\bar{\sigma} \equiv \sigma_{2D} Z_o$ this gives the generic thin film limit results

$$R = O(\bar{\sigma}^2), \; T = 1 - \mathrm{Re}\,\bar{\sigma} + O(\bar{\sigma}^2) \; \text{and} \; A = \mathrm{Re}\,\bar{\sigma} + O(\bar{\sigma}^2). \tag{7}$$

For a single-crystal graphene sample on a SiO$_2$ substrate one has experimentally measured (between *0.5* and *1.2 eV*) [19] a sheet conductivity with a value very close to $\sigma_{2D} = e^2/4\hbar \equiv \alpha\pi/Z_o$ where $\alpha$ is the fine structure constant. This is in accordance with predictions based on interband transitions, within a model with non-interacting massless Dirac fermions, giving an absorption $A = \bar{\sigma} = \alpha\pi \approx 2.3\%$.

The purpose of the next section is to point out yet a third interesting aspect with respect to the light reflection and transmission in an embedded graphene system which is based on finding an optimum sheet conductivity in order to maximize the absorption. This means finding $\bar{\sigma}$ - values of the order of unity; in other words a 2D sheet conductivity being two orders of magnitude $(1/\alpha)$ larger than the one above only giving an absorption $A \sim O(\alpha)$.

### 3. Realization of optimal conditions tuning graphene conductivity

In the previous section we went through the standard argument for obtaining the transmission, reflection and absorption in a thin film. In this section we instead ask ourselves the more general question: what conductivity should the thin film (graphene) have to maximize its absorption at a particular wavelength and possibly for a range of wavelengths? Introducing

$$\Sigma \equiv \frac{Z_3 Z_1}{Z_3 + Z_1} \sigma_{2D} \tag{8}$$

and

$$\Delta \equiv \frac{Z_3 - Z_1}{Z_3 + Z_1}, \qquad (9)$$

we can write for the total absorption

$$A = \frac{2\text{Re}[\Sigma(1+\Delta^*)]}{|1+\Sigma|^2}. \qquad (10)$$

Having graphene surrounded by pure dielectrics, $1+\Delta^*$ can be taken outside the real-part operator so that the absorption is maximum for $\Sigma = 1$. In other words

$$A_{\max} = \frac{Z_3}{Z_3 + Z_1} =_{normal\ incidence} \frac{n_1}{n_3 + n_1}, \qquad (11)$$

given that we can control the graphene conductivity to fulfill

$$\sigma_{2D} Z_o = (\frac{1}{Z_3} + \frac{1}{Z_1}) Z_o =_{normal\ incidence} n_3 + n_1. \qquad (12)$$

In general, as a function of angle of incidence $\theta$, $A_{\theta=0} < A_{\max}(\theta) < A_{\theta=90}$ for $\varepsilon_3 > \varepsilon_1$ and $A_{\theta=90} < A(\theta) < A_{\theta=0}$ for $\varepsilon_1 > \varepsilon_3$ where $A_{\theta=0} = n_1/(n_3 + n_1)$ and $A_{\theta=90} = 1/[1+(\varepsilon_3/\varepsilon_1)\sqrt{(\varepsilon_1 - \varepsilon_o)/(\varepsilon_3 - \varepsilon_o)}]$. Notice that in a symmetric situation, $A_{\max}(\theta) = A_{\theta=0} = A_{\theta=90} = 1/2$, independent of the surrounding media as long as they are identical and that Eq. (12) is fulfilled.

Similarly, we find that the normalized conductivity that provides maximum absorption has to be within the relatively narrow range

$$n_1 + n_3 \leq \overline{\sigma} \leq n_1/\sqrt{1-1/n_1^2} + n_3/\sqrt{1-1/n_3^2} \qquad (13)$$

for typical dielectric materials. The left/right hand side is for normal/glancing incidence. These choices provide optimal circumstances for light absorption over a broad angular range. In other words, if we can change the dielectric properties of graphene so they match the conditions above, i.e. entirely expressed in terms of the optical properties of the surrounding media, we can achieve an absorption which is not only finite but can reach values approaching 100% for cases when $Z_3 >> Z_1$. We notice that if chosing a supporting medium such that this condition is fulfilled, only the on-top medium ($Z_1$) needs matching with the graphene conductivity to achieve the large absorption.

In a symmetric embedding of graphene, $Z_1 = Z_3$, $\Delta$ is zero and we have a 50% maximum absorption independent of the value of the actual surface impedance as long as we can fulfill the condition for the graphene conductivity. Whether this is compatible with what we can do changing the graphene conductivity is the subject of the next section. The results presented above for real dielectrics are valid for all angles of incidence as they are expressed in terms of surface impedances. We also want to point out that with indexes of refraction of the order of unity one needs values of the graphene conductivity which are of the order of $1/\alpha$ in units of $e^2/\hbar$. Such situations have been envisaged both theoretically and experimentally in the literature based on both different doping levels as well as gate voltages [20-23].

**4. How to get maximum absorption out of graphene – simplistic estimate**

We have in Equation (12) identified the conductivity of graphene which maximizes the optical absorption. For illustrative purpose we use the simplest possible representation of the graphene dielectric properties with a conductivity (normalized to the vacuum impedance $Z_o$) of the form [24]

$$\bar{\sigma} = \alpha[\frac{4i\mu}{\hbar(\omega+i/\tau)}g(\mu/k_BT) + \pi\theta(\hbar\omega-2\mu) + i\ln\left|\frac{\hbar\omega-2\mu}{\hbar\omega+2\mu}\right|], \quad (14)$$

.

adding intra- and interband contributions with $\theta$ being the Heaviside function and $g(z) \equiv 1 + \frac{2}{z}\ln(1+e^{-z})$. The conductivity is possible to control by doping and/or an applied bias entering through the chemical potential $\mu$. We introduce $x \equiv 2\mu/\hbar\omega$, $y \equiv E_\tau/\hbar\omega$ ($E_\tau \equiv \hbar/\tau$) and $z \equiv \mu/k_BT$. Equation (14) can then be written as

$$\text{Re}\,\bar{\sigma} = \alpha[\frac{2xy}{1+y^2}g(z) + \pi\theta(1-x)], \quad (15a)$$

and

$$\text{Im}\,\bar{\sigma} = \alpha[\frac{2x}{1+y^2}g(z) + \ln\left|\frac{1-x}{1+x}\right|]. \quad (15b)$$

In Eq. (15a) we see the high-frequency absorption $\alpha\pi$ for *x<1* and small damping (small *y*). This is further illustrated in Figure 2a below which shows $\bar{\sigma}$ and the absorption *A* over four orders of magnitude in frequency (at 300K). Obviously the normalized conductivity in Figure 2a is only of the order 0.1. In order to match $\text{Re}\,\bar{\sigma}$ to a number of the order unity (cf. Eq. (12)) we have to abandon the high-frequency inter-band part since we need a very large *x* and a *y=1* to maximize (15a). In other words this puts us in the low-frequency intra-band part of the response since we need $\mu \gg \hbar\omega \sim E_\tau$. However we have to meet the criterion of a small imaginary part for the conductivity, so $y$ in facts needs to be large ($\hbar\omega \ll E_\tau$). In other words the criteria for realizing a high absorption in graphene are (g(z>>1)=1)

$$\operatorname{Re}\bar{\sigma} \approx \frac{2\alpha x}{y} = 4\alpha \frac{\mu}{E_\tau} \qquad (16a)$$

and

$$\operatorname{Im}\bar{\sigma} \approx \frac{2\alpha x}{y^2} = \operatorname{Re}\bar{\sigma} \cdot \frac{\hbar\omega}{E_\tau} \ll \operatorname{Re}\bar{\sigma}. \qquad (16b)$$

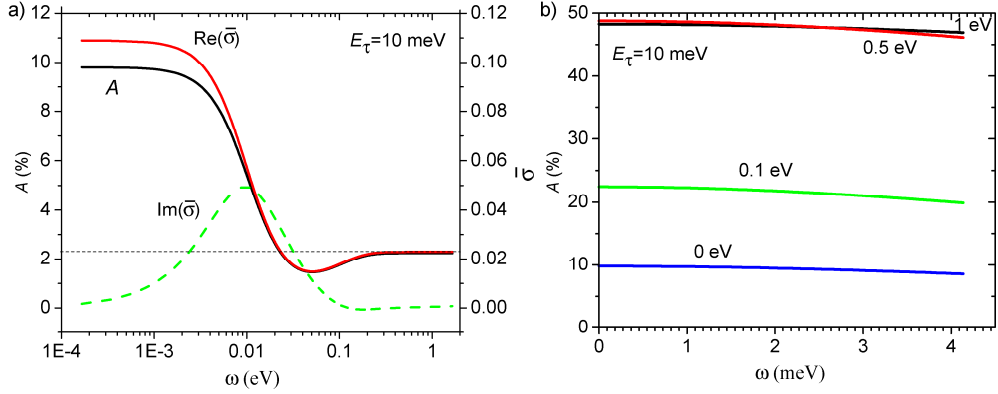

**Figure 2**: a) The real and imaginary parts of the graphene conductivity $\bar{\sigma}$ (Eq.(15)), in units of the vacuum impedance and as a function of the light frequency (in eV) over four orders of magnitude (at 300K). The damping is assumed to be 10 meV and the chemical potential ⍰ is set to zero. For vacuum on either side of the graphene sheet this gives the absorption shown, which approaches the optical limit of 2.3% for higher frequencies. Notice how the real part of the conductivity accounts for the major part of the absorption. b) The strong dependence of the maximum absorption on chemical potential µ for a graphene sheet in vacuum is shown, approaching the symmetric limit of 50% absorption for higher values of the chemical potential, almost independent of the frequency in the chosen range.

Thus neglecting the imaginary part and inserting the real part in Eq. (16a) into Eqs. (8-10) for the absorption we can write in the symmetric situation

$$A(\mu) = \frac{2\bar{\mu}}{(1+\bar{\mu})^2},  \qquad (17)$$

introducing a normalized chemical potential $\bar{\mu} \equiv \mu/\mu_o$. This equation contains the main message of this paper. By carefully controlling the chemical potential through doping or gating, adjusting with respect to a characteristic chemical potential $\mu_o$, we can control the level of absorption in an embedded sheet of graphene between 0 and 50%. This range can be extended to 100% provided we use an asymmetrical embedding of dielectrics or a reflective support [10]. The characteristic chemical potential $\mu_o$ is given from our analysis as

$$\mu_o = \frac{Z_o}{2\alpha Z_1} E_\tau =_{normal\ incidence} \frac{n_1}{2\alpha} E_\tau. \qquad (18)$$

With typical values of the refractive index of the surroundings of the order of unity we get a characteristic chemical potential of the order of $E_\tau$ divided by the fine-structure constant. For a ballistic transport of several hundreds of nanometers [16], which puts $E_\tau$ in the range of *1-10* meV (lifetimes in the range of *0.4* to *4p*s), we get a chemical potential in the range of *0.1-1* eV. This is well within accessible values matching typical surroundings like Si. For illustrative purposes we show in Figure 3 below how the absorption *A* varies as a function of the chemical potential μ; for two different frequencies *100* and *900* GHz, two different dampings *1* and *10* meV and a number of different combinations of the dielectric media surrounding the graphene sheet.

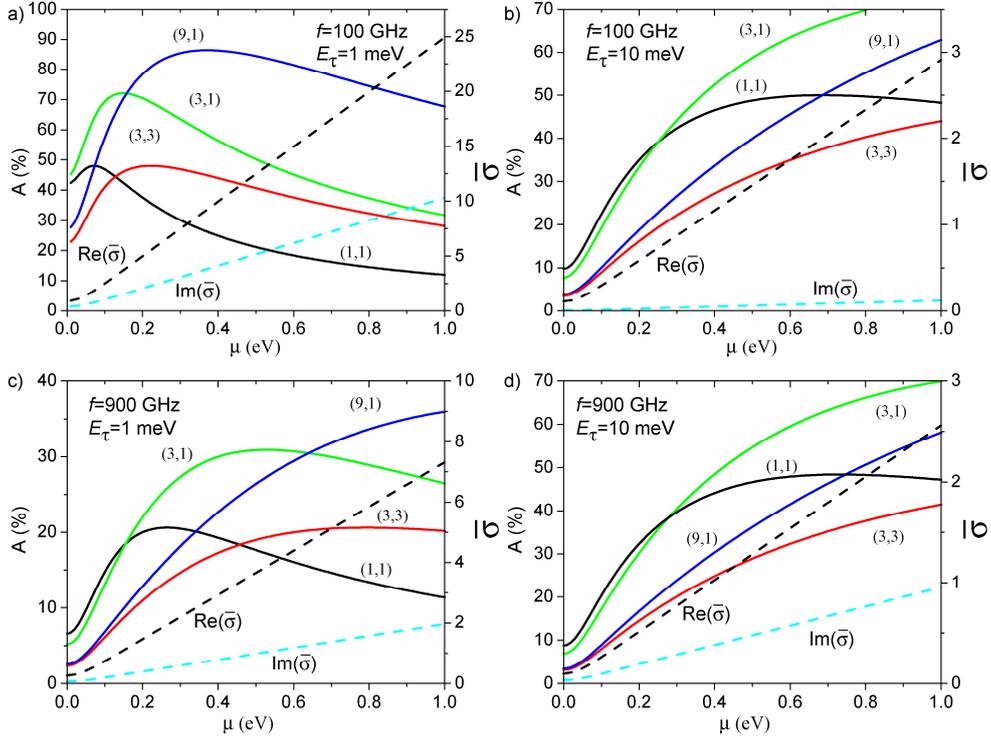

**Figure 3.** Absorption *A*, real and imaginary parts of $\bar{\sigma}$ as a function of the chemical potential, for different damping $E_\tau$ (1 and 10 meV) and light frequencies *f* (100 and 900 GHz). The large absorption results from our maximizing procedure with respect to different dielectric surroundings indicated as ($n_1$, $n_3$). Notice the exceptionally large possible values for the absorption when we have a high dielectric contrast of the embedding.

The only experiments we are aware coming close to probe the suggested behaviour in this paper as shown in Figure 3 are the ones in [25-27]. However further experimental work is needed in order to utilize the possible implications and applications of our Eq. (17).

## 5. Concluding remarks

We have shown in Figure 3 above the absorption *A* as a function of the different relevant physical parameters. We find that for frequencies in the upper GHz/lower THz range it should be possible to match graphene to standard dielectrics reflectors in order to achieve 50-100% absorption of all incoming radiation in a one atom thick material simply by changing the chemical potential by means of voltage bias and/or doping. This can have interesting applications such as modulators, filters and alike [28] as we will further explore awaiting an experimental verification of the predictions made in this paper. Before that we should also address any possible, and potentially useful, effects coming through the interplay between chemical potential and temperature. We most likely need to go beyond a constant relaxation time approximation using a frequency-dependent scattering rate. Finally we notice that for a situation with an appropriate meta-material present one can achieve situations where the surface impedance of the surrounding becomes mainly imaginary and this interchanges the role of real and imaginary parts of the conductivity in Eq. (14); a phenomenon which needs further exploration.

**Acknowledgements.** One of us (SPA) acknowledges a grant from the Chalmers' *Area of Advance in Nanoscience and Nanotechnology* as well as support from the Swedish Foundation for Strategic Research via the metamaterial project SSF RMA08. We had enlightening discussions with Andreas Isacsson, Weihua Wang and Jari Kinaret as well as fruitful input from A.C. Ferrari.